%% file: msv3.tex
\documentclass[12pt,preprint,useAMS,usenatbib,usegraphicx,english]{aastex}

\usepackage[T1]{fontenc}
\usepackage[utf8]{inputenc}
\usepackage{babel}

\shortauthors{Castro & Gizis}
\shorttitle{WISEP J060738.65+242953.4}
\begin{document}

\title{DISCOVERY OF A LATE L DWARF: WISEP J060738.65+242953.4}

\author{Philip J. Castro\altaffilmark{1} and John E. Gizis\altaffilmark{1}}
\affil{Department of Physics and Astronomy, University of Delaware, Newark, DE 19716; pcastro@udel.edu, gizis@udel.edu}

\begin{abstract}
We discover a late-type L dwarf, WISEP J060738.65+242953.4 (W0607+2429), by comparing the Wide-field Infrared Survey 
Explorer (WISE) preliminary data release to the Two Micron All Sky Survey (2MASS) in search of
high proper motion objects ($\gtrsim$$0\farcs3$ yr$^{-1}$). W0607+2429 was found to have a proper motion 
of $0.57\pm0\farcs02$ yr$^{-1}$. Based on colors and color-color diagrams using 2MASS and Sloan Digital 
Sky Survey (SDSS) photometry, we estimate the spectral type (optical) to be L8 within a spectral sub-type. Based 
on the spectral type estimated we find W0607+2429 to have a distance of 7.8$^{+1.4}_{-1.2}$ pc, making it one of only 
four very-late L dwarfs within 10 pc, and the third closest L dwarf overall. This close L/T transition 
dwarf will play a pivotal role in resolving outstanding issues of condensate clouds of low temperature atmospheres.
\end{abstract}


\section{INTRODUCTION}
The Wide-field Infrared Survey Explorer (WISE) mission is an all-sky survey, whose bands are centered
on wavelengths 3.4$\mu$ ($W1$), 4.6$\mu$ ($W2$), 12$\mu$ ($W3$), and 22$\mu$ ($W4$), 
achieving 5$\sigma$ point source sensitivies.
One of the main scientific goals of the WISE mission is to detect cool brown dwarfs (BDs), 
ranging from T dwarfs to the evasive Y dwarfs \citep{Wrightetal2010,Mainzeretal2011}.
It accomplishes this by observing at wavelengths where the spectral energy distribution of late T dwarfs 
and Y dwarfs peak \citep{Wrightetal2010}. The WISE preliminary release has
yielded multiple late T dwarf discoveries \citep{Burgasseretal2011b,Mainzeretal2011,Wrightetal2011}.
As an all-sky survey, WISE provides an ideal platform, in conjunction with other
all-sky surveys such as 2MASS, to study the proper motion of brown dwarfs by creating an all-sky multi-epoch survey.
By comparing WISE to 2MASS, with similar photometric bands, and a sizable difference in epochs, $\sim$ 10 yr, these 
two all-sky surveys provide an ideal setup to find brown dwarfs with large proper motion (nearby brown dwarfs). 
Multi-epoch searches using WISE have already produced numerous brown dwarf discoveries 
\citep{Aberasturietal2011,Liuetal2011,Loutreletal2011,Gizisetal2011a,Gizisetal2011b,Scholzetal2011}.

Late L dwarfs are characterized by very red near-infrared colors ($J-K_{\rm s}\sim2$), H$_{2}$O absorption,
and CO absorption in the $K$ band. Early T dwarfs have a reversal of near-infrared colors to 
blue ($J-K_{\rm s}\sim0$), brightening of the $J$ band \citep{Dahnetal2002}, weakening of CO absorption 
and the strengthening of CH$_{4}$ (the onset of CO to CH$_{4}$ conversion) and H$_{2}$O absorption, 
where the unambigous detection of CH$_{4}$ at the $H$ and $K$ bands is the defining 
characteristic of T dwarfs \citep{Kirkpatrick2005}. 
This L/T transition occurs over a small temperature range of $\sim$200-300 K \citep{Kirkpatrick2005},
and is believed to be caused by the depletion of condensate clouds, where the driving mechanism for the depletion
is inadequately explained by current cloud models \citep{Burgasseretal2011a}.
The bluer $J-K$ and the brightening 
of the $J$ band at the L/T transition can be explained by decreasing cloudiness. A mechanism 
suggested for the L/T dwarf spectral type transition is the appearance of relatively cloud free regions 
across the disk of transition L/T dwarfs as they cool \citep{Marleyetal2010}. The complex dynamic behavior of 
condensate clouds of low temperature atmospheres at the L/T transition is one of the 
leading problems in brown dwarf astrophysics today \citep{Burgasseretal2011a}. 

There are dozens of known very-late L dwarfs at the L/T transition.
There are 26 L7-L8 dwarfs with optical (opt) classification and
24 L7-L9.5 dwarfs with solely near-infrared (NIR) classification listed in the Dwarf
Archives as of 14 February 2011 \citep{Gelinoetal2009}, and four additional L7-L8
dwarfs from \citet{Schmidtetal2010}.
However, there are only three very-late L dwarfs within 10 pc. 
The L8 (opt) dwarf DENIS-P J0255-4700 \citep{Martinetal1999} at $4.97\pm0.10$ pc \citep{Costaetal2006}, 
the recently discovered L7.5 (NIR) dwarf WISEP J180026.60+013453.1 at 8.8$\pm$1.0 pc \citep{Gizisetal2011a}, 
and the L8 (opt) dwarf 2MASS J02572581-3105523 \citep{Kirkpatricketal2008} at 9.7$\pm$1.3 pc \citep{Looperetal2008b}.
Clearly very-late L dwarfs within 10 pc are rare. 
These close L/T transition BDs are fundamental in providing observational constraints to 
understanding the low temperature atmospheres of these objects.

We present the discovery of a late L dwarf, WISEP J060738.65+242953.4 (W0607+2429), as part of a continued effort to 
discover brown dwarfs by their high proper motion between 2MASS and WISE \citep{Gizisetal2011a,Gizisetal2011b}.
In section 2 we present our analysis; the discovery of W0607+2429, determine the proper motion, estimate the spectral type 
based on colors and color-color diagrams, distance, and other physical properties. In section 3 our conclusions will 
summarize our findings and discuss future work.

\section{ANALYSIS}
\subsection{Discovery}
We used the same criteria to search for high proper motion objects as \citet{Gizisetal2011b},
but extended the search to red colors.
We searched for WISE sources that had detections at $W1$ (3.4$\mu$), $W2$ (4.6$\mu$), and $W3$ (12$\mu$), 
no 2MASS counterpart within $3^{\prime \prime}$, and red colors $W1-W2>0.3$. WISE and 2MASS
images were used to create finder charts to visually search for high proper motion candidates.
WISEP J060738.65+242953.4 (W0607+2429) was found to have a separation of $\approx7^{\prime \prime}$ from a 2MASS source to 
the northeast, 2MASSW J06073908+2429574. The WISE source shows colors that are red, $W1-W2=0.60\pm0.05$, 
consistent with that of a late L dwarf/early T dwarf \citep{Mainzeretal2011}, where the 2MASS source has
red colors that are consistent with an L dwarf \citep{Kirkpatricketal2000}, $J-H=1.18\pm0.05$ and $H-K_{\rm s}=0.57\pm0.05$. 
SDSS J060738.79+242954.4 (DR7) was found between the 2MASS and WISE positions at an intermediate epoch,
and was recognized as having very red colors, $i-z=3.08\pm0.04$, indicative of a late L dwarf \citep{Schmidtetal2010},
see Figure 1 bottom right image.
We positively identify the 2MASS and the Sloan Digital Sky Survey (SDSS) 
source as W0607+2429 at their respective epochs. With a high proper motion indicating a nearby object and red 
colors in 2MASS, SDSS, and WISE indicating a late spectral type, we confidently claim the detection of a 
nearby ultracool dwarf. A finder chart for W0607+2429 showing a clear linear sequence of positions at the epoch of 2MASS, 
SDSS, and WISE is shown in Figure 1.

\begin{figure}
\caption{
Finder chart showing the proper motion of W0607+2429 from the 2MASS $K_{\rm s}$ band image (top left) to the 
SDSS $z$ band image (top right) (DR8, run 6585, rerun 301) to the WISE $W1$ image (bottom left). 
The three circles in each image show, from top left to bottom right, the position of W0607+2429 at the 2MASS, 
SDSS, and WISE positions, respectively. The bottom right image is an RGB image of SDSS, where the $z$ band is
red, the $i$ band is green, and the $r$ band is blue. The prominent red source in the SDSS RGB image, 
W0607+2429, is unmistakably a late-type star. North is up and east is to the left.
}
\includegraphics[width=6.5in]{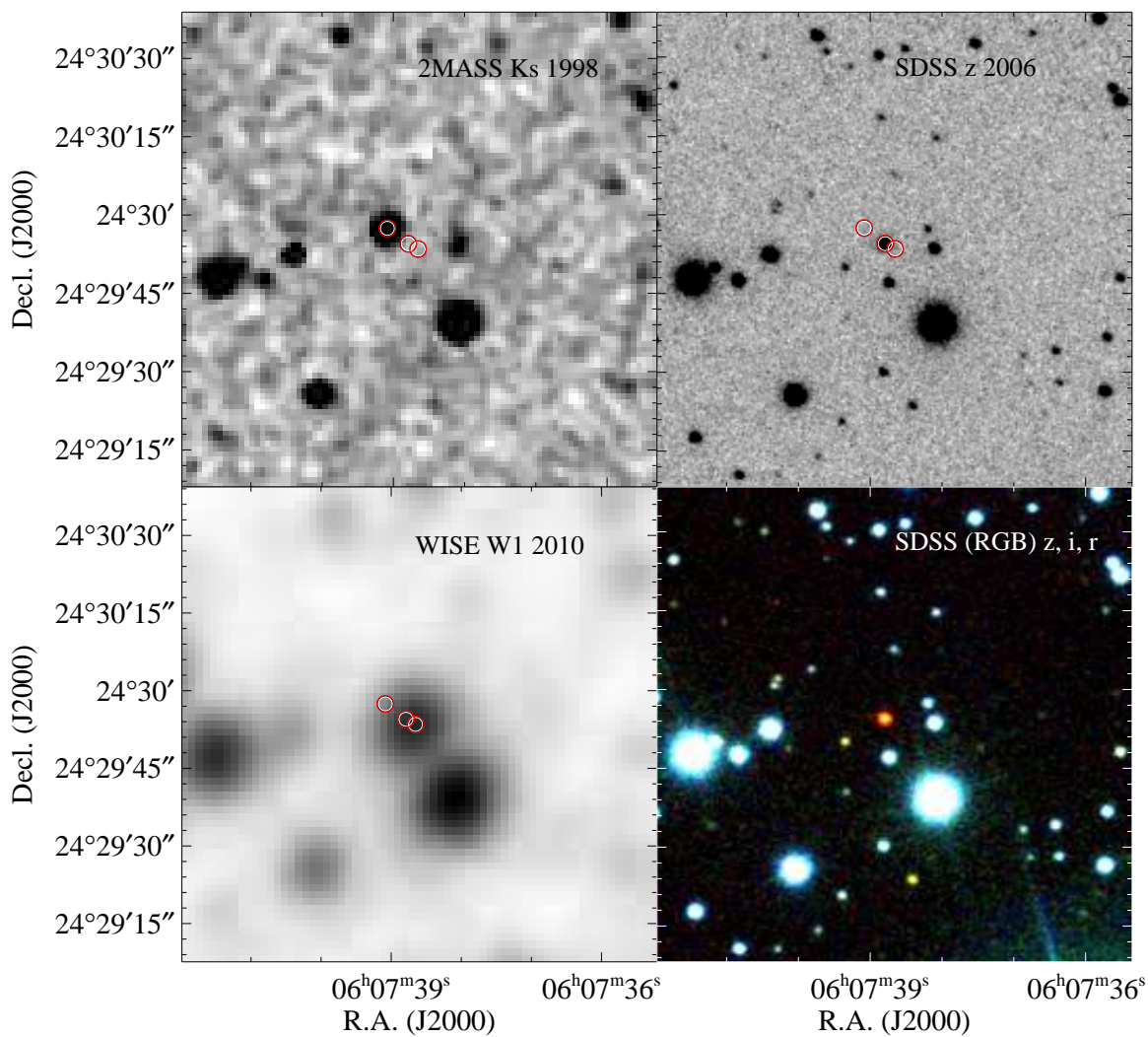}
\end{figure}

\subsection{Proper Motion}
We calculate the difference in position of W0607+2429 between the 2MASS, SDSS, and WISE epochs based on reference 
stars within 5$^{\prime}$, with the uncertainty in position based on the uncertainties in the 2MASS, SDSS, and 
WISE catalogs. We determine the proper motion of W0607+2429 by using a linear least-squares fit to the relative 
position at the 2MASS, SDSS, and WISE epochs, as shown in Figure 2. 
We find a proper motion of $\mu_{\alpha}$cos($\delta$)$=-0.47\pm0\farcs01$ yr$^{-1}$ 
and $\mu _{\delta}=-0.33\pm0\farcs02$ yr$^{-1}$, with total motion $0.57\pm0\farcs02$ yr$^{-1}$.
We corrected for the parallactic motion of the 2MASS, SDSS, and WISE positions using NOVAS V3.0 
software \citep{Kaplanetal2009} based on the estimated distance (see Section 2.4) of W0607+2429.
In the WISE preliminary release source catalog a pipeline processing error resulted in a declination bias of $0\farcs5$,
to account for the declination bias the actual declination errors of all WISE sources were inflated by adding
a $0\farcs5$ error term in quadrature.
However, it was discovered that this pipeline processing error affected WISE sources fainter than $W1>13.0$.
We restricted WISE sources to $W1<13.0$ and W0607+2429 ($W1<13.0$) is not affected, we removed this $0\farcs5$
error term from the reported declination error to determine the actual error in calculating the proper motion.
For more details see the Explanatory Supplement to the WISE Preliminary Data Release
Products\footnote{http://wise2.ipac.caltech.edu/docs/release/prelim/expsup/sec6\_5.html}.
We use the astrometry and photometry from SDSS DR7 \citep{Abazajianetal2009} rather than DR8 \citep{Aiharaetal2011} 
due to astrometric errors associated with DR8; we use the DR8 image in Figure 1. We note that the astrometry 
is different by 50 mas for W0607+2429, and the photometry in the $i$ and $z$ band are almost identical for 
W0607+2429, between DR7 and DR8. For additional information regarding the astrometric
errors in DR8 refer to SDSS III\footnote{http://www.sdss3.org/dr8/algorithms/astrometry.php\#caveats}.

\begin{figure}
\caption{
Best fit line determining proper motion based on the relative 2MASS, SDSS (DR7), and WISE positions
of W0607+2429 for right ascension (left) and declination (right).
}
\includegraphics[width=3in]{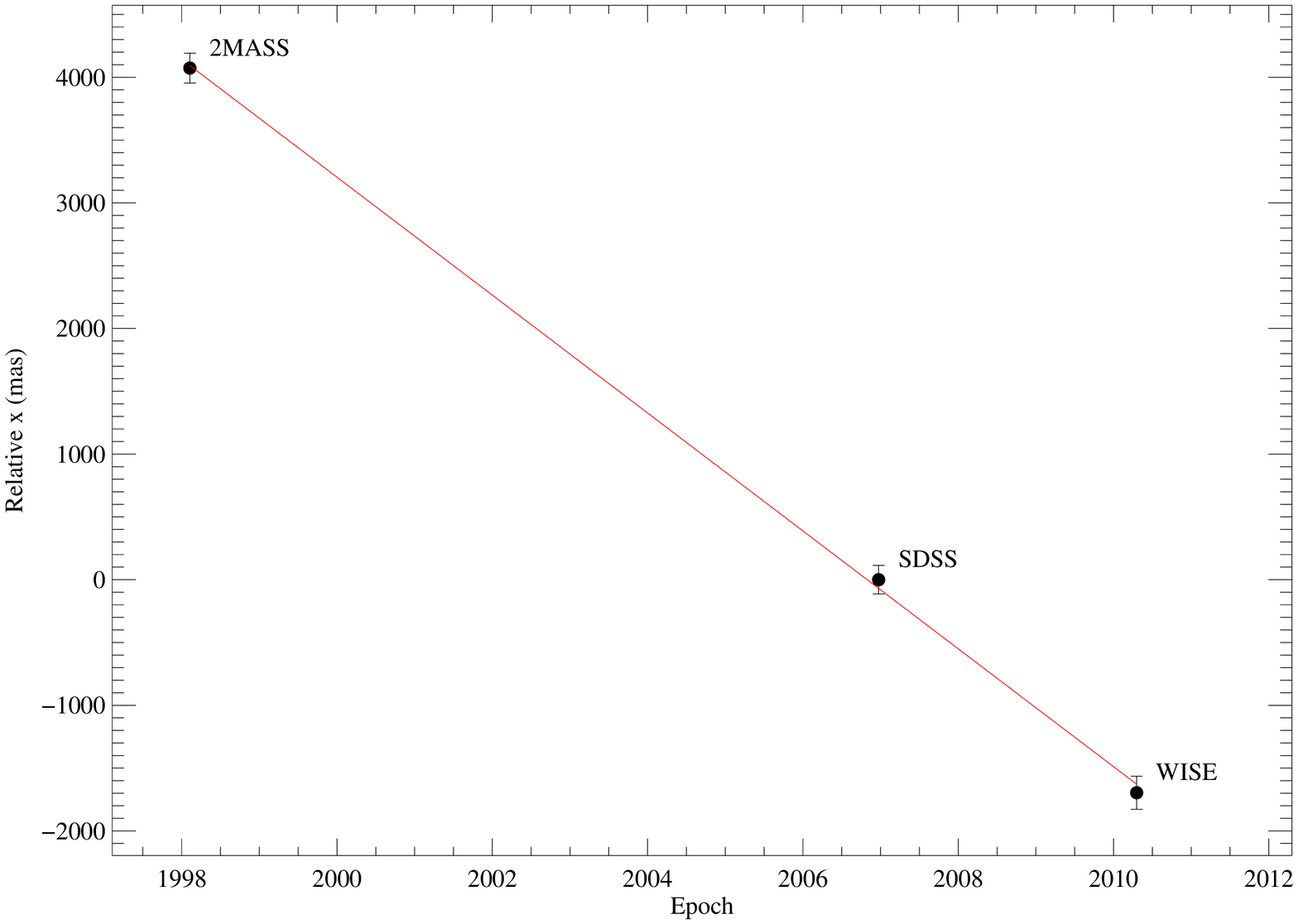}
\includegraphics[width=3in]{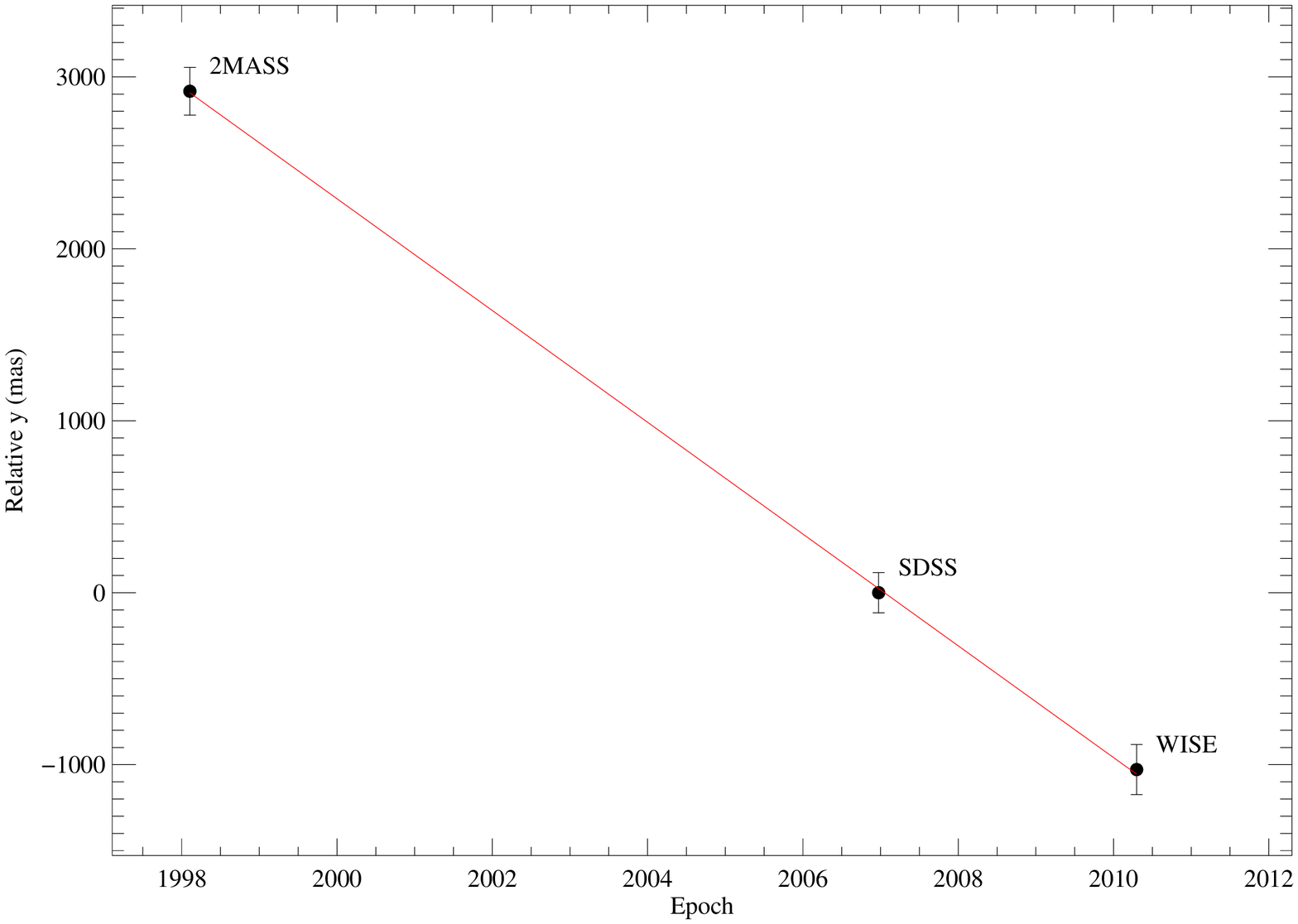}
\end{figure}

\subsection{Spectral Type Estimate}
We reproduce the color vs spectral type plots for $i-z$ and $i-J$ from \citet{Schmidtetal2010}, 
and overplot W0607+2429, see Figure 3. The $i-z$ and $i-J$ colors are relatively good predictors of spectral 
type \citep{Schmidtetal2010}. It is clear from Figure 3 that W0607+2429 is a late L dwarf.

\begin{figure}
\caption{
Color vs spectral type for L dwarfs from \citet{Schmidtetal2010} with W0607+2429 overplotted.
The diamonds show the mean values of color for each spectral type from \citet{Schmidtetal2010} with 
the error bars showing the standard deviation (the standard deviation reflects the intrinsic scatter 
in each spectral type). The red circle is W0607+2429, whose position in color 
space is consistent with that of a late L dwarf.
}
\includegraphics[width=6in]{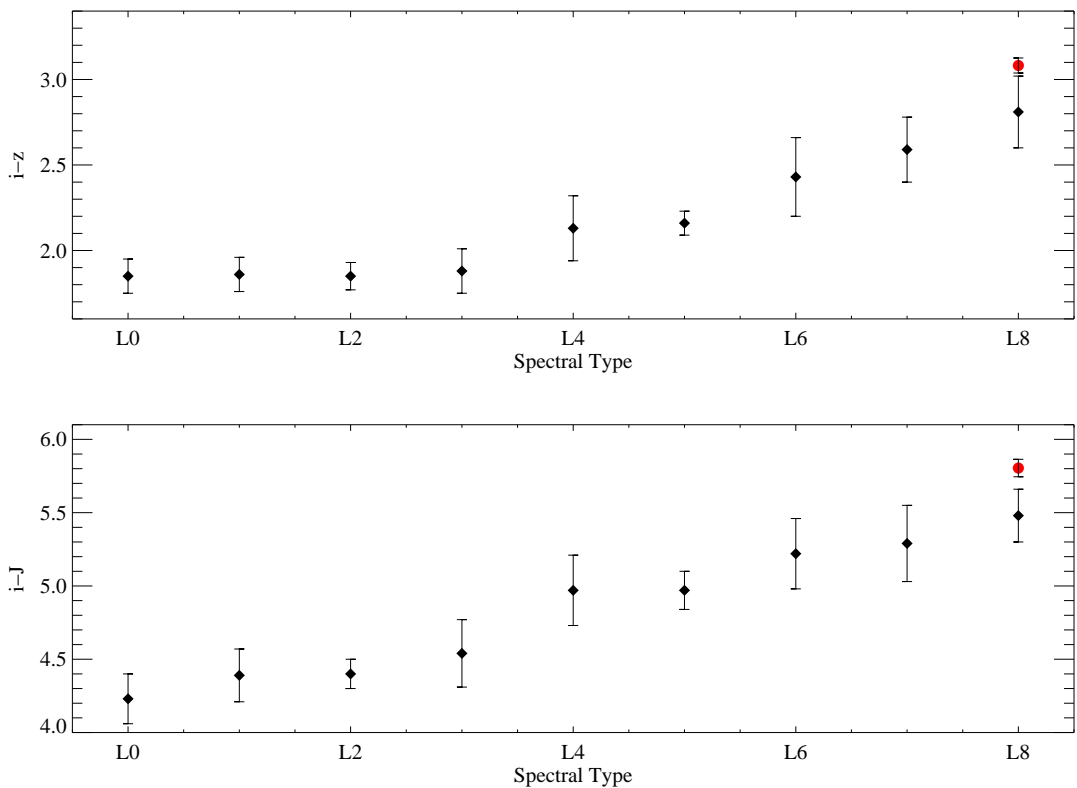}
\end{figure}

We reproduce color-color diagrams from \citet{Schmidtetal2010} using their Table 1 data and one additional
L8 dwarf from their Table 5. Figure 4 shows four color-color diagrams, where all four show W0607+2429 consistently 
lies in the color-color space that is the locus of L8 dwarfs.
Based on the colors of W0607+2429, with confidence we estimate the spectral type as L8 within a spectral sub-type.
We note that W0607+2429 has very simlar colors to the L8 dwarf 2MASSW J1632291+190441 \citep{Kirkpatricketal1999} 
at $\approx15$ pc \citep{Dahnetal2002}, with $i-z=3.11\pm0.20$,
$i-J=5.82\pm0.23$, $z-J=2.71\pm0.11$, and $J-K_{\rm s}=1.86\pm0.12$.
W0607+2429 has colors that are too red to be considerd a T0 dwarf, with $J-W_{2}=3.27\pm0.05$ and 
$H-W_{2}=2.09\pm0.05$ \citep{Mainzeretal2011}, $J-K_{\rm s}\not\sim0$ \citep{Kirkpatrick2005},
and 2MASS colors that place W0607+2429 far from the T dwarf locus in a $J-H$, $H-K_{\rm s}$ 
color-color diagram \citep{Kirkpatricketal2000}.

\begin{figure}
\caption{
Color-color diagrams reproduced with data from \citet{Schmidtetal2010}.
The black boxes are early L dwarfs, L0-L4, and the larger colored boxes
are L5-L8 dwarfs. L5 dwarfs are magenta, L6 dwarfs are yellow, L7 dwarfs are blue,
and L8 dwarfs are green. The red circle is W0607+2429, which lies 
at the locus of L8 dwarfs in color-color space.
}
\includegraphics[width=6.5in]{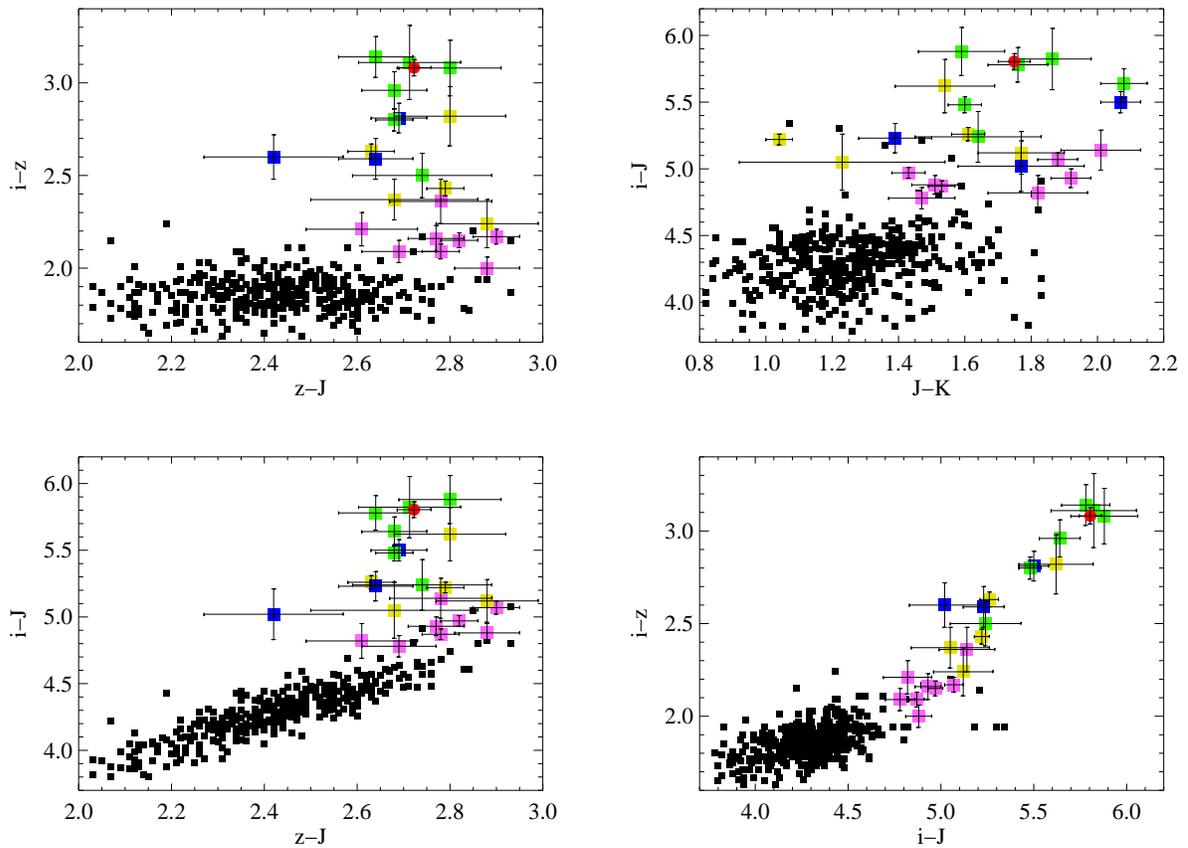}
\end{figure}

\subsection{Distance}
We estimate the distance by using the spectral type-absolute magnitude relationships 
from \citet{Looperetal2008a} for 2MASS photometry and \cite{Schmidtetal2010} for SDSS photometry. 
We find distance estimates of $7.9^{+1.2}_{-1.1}$ pc from 2MASS $J$ photometry, $7.9^{+1.2}_{-1.1}$ pc 
from 2MASS $H$ photometry, $8.4^{+1.3}_{-1.1}$ pc from 2MASS $K_{\rm s}$ photometry, and $7.2^{+1.6}_{-1.3}$ pc 
from SDSS $i$ and $z$ photometry, where the uncertainty in the distance estimates comes from the uncertainty 
in the photometry and the RMS from the spectral type-absolute magnitude relationships. 
The mean of these estimates provides a distance of $7.8^{+1.4}_{-1.2}$ pc, 
assuming no binarity. 
This distance estimate places it as the third closest L dwarf, after 
the L8 dwarf DENIS-P J0255-4700 \citep{Martinetal1999} at $4.97\pm0.10$ pc \citep{Costaetal2006}
and the L5 dwarf 2MASSW J1507476-162738 \citep{Reidetal2000} at 7.33$\pm$0.03 \citep{Dahnetal2002}.
The uncertainty in distance gives W0607+2429 a range of the second closest to the seventh closest L dwarf, refer
to \citet{Gizisetal2011a} for a discussion of L dwarfs within 10 pc.
This proximity of W0607+2429 brings the number of very-late L dwarfs within 10 pc from three to four. 
Trigonometric parallax measurements are needed for a more reliable distance estimate.

\subsection{Other Physical Properties}
W0607+2429 has a tangential velocity of 21$^{+4}_{-3}$ km s$^{-1}$, within range of 
transverse motions for other L8 dwarfs from \citet{Fahertyetal2009}, who quote a median value 
of 25 km s$^{-1}$ and a dispersion of 19 km s$^{-1}$.
This $v_{\rm tan}$ is consistent with that expected for a member of the Galactic 
thin disk \citep{Fahertyetal2009}.
Spectral type-effective temperature \citep{Looperetal2008a} and spectral type-absolute 
bolometric magnitude \citep{Burgasser2007} relationships give a $T_{\rm eff}=1460\pm90$ K and 
a log $L/L_{\odot}=-4.56\pm0.09$, where the uncertainty in $T_{\rm eff}$ comes from the RMS in the 
spectral type-effective temperature relation and the uncertainty in log $L/L_{\odot}$ is from the RMS 
in the spectral type-absolute bolometric magnitude relation.
Based on these physical properties, theoretical isochrones from \citet{Baraffeetal2003} give a range 
of 0.5 Gyr and 0.03 M$_{\odot}$ to 10 Gyr and 0.072 M$_{\odot}$, W0607+2429 is in the substellar regime,
as are all of the latest L dwarfs \citep{Kirkpatrick2005}.

Field binaries are primarily equal brightness/mass systems in tightly bound orbits ($<$ 20 AU), 
where the separation of binary systems peaks at $<$ 10 AU \citep{Allen2007,Burgasseretal2007}. 
A secondary to W0607+2429 of equal or earlier spectral type ($\lesssim$L8) would have been detected 
at $\gtrsim$ 8 AU based on the FWHM ($\approx1^{\prime \prime}$) of SDSS in the $i$ and $z$ band. 
If W0607+2429 was an unresolved binary system, for example, consisting of two L8 dwarfs, it would push 
the distance estimate out to 11.1 pc. 
The highest resolution imaging/spectroscopy is warranted to search for a companion to W0607+2429. The 
sensitivity of current imaging surveys begins to fall off at separations of $\lesssim3-4$ AU, where 
there is a model predicted frequency peak of binarity \citep{Allen2007}. Rare nearby L dwarfs 
\citep{Gizisetal2011a} such as W0607+2429, if found to have companions, will help to fill this void. 
A summary of characteristics for W0607+2429 is found in Table 1.

\input{tab1.dat}

\section{CONCLUSIONS}
We have discovered a high proper motion late L dwarf, WISEP J060738.65+242953.4 (W0607+2429), with a 
proper motion of $0.57\pm0\farcs02$ yr$^{-1}$ and an estimated spectral type (optical) of L8 based on 
its colors. We estimate a distance of 7.8$^{+1.4}_{-1.2}$ pc based on this spectral type, placing W0607+2429 
as the third closest L dwarf, and one of only four very-late L dwarfs within 10 pc.

Follow-up spectroscopy is necessary to confirm the spectral type of W0607+2429, parallax measurements are needed
to determine the distance with more confidence, and the highest resolution imaging/spectroscopy is warranted 
to determine binarity. Observations to determine the photometric variability and polarization of W0607+2429 
will address theories regarding the inhomogeneity of cloud cover and the color change across 
the L/T transition \citep{Marleyetal2010}.
Improving these inadequate models of L/T transition dwarf atmospheres has implications beyond brown dwarfs, such 
as hot exoplanets (HR 8799b) that are analogs to L and T dwarfs \citep{Fortney2005,Currieetal2011}. W0607+2429 will 
serve as a fundamental testbed to further resolve outstanding issues regarding the L/T transition.

\section{ACKNOWLEDGMENTS}
We thank the Annie Jump Cannon Fund at the University of Delaware for support.
This research has benefitted from the M, L, and T dwarf compendium housed at DwarfArchives.org and 
maintained by Chris Gelino, Davy Kirkpatrick, and Adam Burgasser. 
This publication makes use of data products
from the Two Micron All Sky Survey, which is a joint
project of the University of Massachusetts and the Infrared Processing
and Analysis Center/California Institute of Technology,
funded by the National Aeronautics and Space Administration
and the National Science Foundation.
Funding for SDSS-III has been provided by the Alfred P. Sloan Foundation, the Participating Institutions, 
the National Science Foundation, and the U.S. Department of Energy. SDSS-III is managed by the Astrophysical 
Research Consortium for the Participating Institutions of the SDSS-III Collaboration including the University 
of Arizona, the Brazilian Participation Group, Brookhaven National Laboratory, University of Cambridge, 
University of Florida, the French Participation Group, the German Participation Group, the Instituto de 
Astrofisica de Canarias, the Michigan State/Notre Dame/JINA Participation Group, Johns Hopkins University, 
Lawrence Berkeley National Laboratory, Max Planck Institute for Astrophysics, New Mexico State University, 
New York University, Ohio State University, Pennsylvania State University, University of Portsmouth, 
Princeton University, the Spanish Participation Group, University of Tokyo, University of Utah, 
Vanderbilt University, University of Virginia, University of Washington, and Yale University.
This publication makes use of data products from the Wide-field Infrared Survey Explorer, which is 
a joint project of the University of California, Los Angeles, and the Jet Propulsion Laboratory/California 
Institute of Technology, funded by the National Aeronautics and Space Administration.

\bibliographystyle{apj}
\bibliography{bibliography}

\label{lastpage}

\end{document}

%% file: tab1.dat
\begin{deluxetable}{cc}
\tablecaption{Parameters of WISEP J060738.65+242953.4}
\tablewidth{0pt}
\tablehead{
\colhead{Parameters} & \colhead{WISEP J060738.65+242953.4}
}
\startdata
WISE R.A. (J2000) &06:07:38.65\\
WISE Decl. (J2000) &+24:29:53.5\\
WISE Epoch\tablenotemark{a} &2010.30\\
SDSS R.A. (J2000) &06:07:38.79\\
SDSS Decl. (J2000) &+24:29:54.5\\
SDSS Epoch &2006.98\\
SDSS Data Release/Run/Rerun &DR7/6586/648\\
2MASS R.A. (J2000) &06:07:39.08\\
2MASS Decl. (J2000) &+24:29:57.5\\
2MASS Epoch &1998.11\\
$i-z$ [mag] &3.08$\pm$0.04\\
$i-J$ [mag] &5.80$\pm$0.06\\
$z-J$ [mag] &2.72$\pm$0.04\\
$J-K_{\rm s}$ [mag] &1.75$\pm$0.05\\
WISE $W1$ [mag] &11.55$\pm$0.03
\\
WISE $W2$ [mag] &10.95$\pm$0.02
\\
WISE $W3$ [mag] & 9.93$\pm$0.05
\\
WISE $W4$ [mag] &$>$ 8.54\\
SDSS $i$ [mag] &20.02$\pm$0.03
\\
SDSS $z$ [mag] &16.94$\pm$0.01
\\
2MASS $J$ [mag] &14.22$\pm$0.03
\\
2MASS $H$ [mag] &13.04$\pm$0.03
\\
2MASS $K_{\rm s}$ [mag] &12.47$\pm$0.02
\\
Spectral Type (optical est.) &L8\\
$\mu_{\alpha}$cos($\delta$) (mas yr$^{-1}$) &-470$\pm$10\\
$\mu_{\delta}$ (mas yr$^{-1}$) &-330$\pm$20\\
Distance (pc) &7.8$^{+1.4}_{-1.2}$\\
${v}_{\rm tan}$ (km s$^{-1}$) &21$^{+4}_{-3}$\\
$T_{\rm eff}$ (K) &1460$\pm$90\\
log $L/L_{\odot}$ &-4.56$\pm$0.09\\
\enddata
\tablenotetext{a}{The WISE epoch is the average of the first and second run.}
\end{deluxetable}